\documentclass[aps,superscriptaddress,twocolumn,twoside,floatfix,pra,a4paper]{revtex4-1} 

 \usepackage{lipsum}
\usepackage{times}
\usepackage{epsfig}
\usepackage{amsfonts}
\usepackage{amsmath}
\usepackage{amssymb,amsthm}
\usepackage{xcolor,colortbl}
\usepackage{multirow}
\usepackage{braket}
\usepackage{latexsym}
\usepackage{amsfonts}
\usepackage{mathrsfs}
\usepackage{natbib}
\usepackage{verbatim}
\usepackage{gensymb}
\usepackage{caption}
\usepackage{caption}
\usepackage{subcaption}
\usepackage{ragged2e}
\DeclareCaptionJustification{justified}{\justifying}
\usepackage{blkarray}
\usepackage{graphicx}

\usepackage[colorlinks=true,linkcolor=blue,citecolor=blue,urlcolor=blue]{hyperref}
\allowdisplaybreaks

\begin{document}

\title{{Zero-Error Nash Equilibrium: Harnessing Nonlocal Correlation in Incomplete Information Games}}

\author{Ambuj}
\affiliation{Indian Institute of Technology, Jodhpur-342030, India}

\author{Tushar}
\affiliation{Indian Institute of Technology, Jodhpur-342030, India}

\author{Siddharth R. Pandey}
\affiliation{Morito Institute of Global Higher Education, Hiroshima University, 1-1-1 Kagamiyama, Higashi-Hiroshima City Hiroshima, Japan 739-8524.}

\author{Ram Krishna Patra}
\affiliation{Department of Physics of Complex Systems, S. N. Bose National Center for Basic Sciences, Block JD, Sector III, Salt Lake, Kolkata 700106, India.}

\author{
Anandamay Das Bhowmik}
\affiliation{School of Physics, IISER Thiruvananthapuram, Vithura, Kerala 695551, India.}

\author{Kuntal Som}
\affiliation{Indian Institute of Technology, Jodhpur-342030, India}

\author{Amit Mukherjee}
\affiliation{Indian Institute of Technology, Jodhpur-342030, India}

\begin{abstract}
Claude Shannon’s zero-error communication paradigm reshaped our understanding of fault-tolerant information transfer. Here, we adapt this notion into game theory with incomplete information. We ask: can players with private information coordinate on a Nash equilibrium with zero probability of error? We identify Bayesian games in which such coordination is impossible classically, yet achievable by harnessing Bell nonlocal correlations. We formalize this requirement as \textit{zero-error Nash equilibrium coordination}, establishing a new bridge between information theory, game theory, and quantum nonlocality. Furthermore, we construct a tripartite Bayesian game that admits zero-error Nash equilibrium coordination with genuine entanglement, and a two-player game where a stronger notion of coordination can be achieved using every two-qubit pure entangled state except the maximally one. Crucially, the advantage persists under experimentally relevant noise, demonstrating nonlocality as a robust resource for near-zero error decision-making under uncertainty.
\end{abstract}

\maketitle
\section{Introduction}
Foundational concepts like Shannon’s zero-error capacity \cite{Shannon1956} and correlated Nash equilibrium in Bayesian games \cite{Nash1950,Nash1951,Harsanyi1967i,Harsanyi1968ii,Harsanyi1968iii,Aumann1974} have profoundly shaped our understanding of communication and decision-making. Yet, when reconsidered from a new perspective, these classical ideas can reveal surprising synergies. This paper presents one such union: 
{we ask what happens when players in a Bayesian game are required to coordinate on their intended strategic outcome with zero probability of error}
—a condition inspired by Shannon’s zero-error communication paradigm. Though developed independently, these ideas exhibit deep connections when examined within a unified framework. 
{Game theory typically focuses on the existence of a Nash equilibrium for a game, rather than on guaranteeing that players actually reach it in every play. Motivated by high-stakes scenarios where even a single deviation can be catastrophic, our framework instead requires that a Nash equilibrium be achieved without error in every round of game play.}
We show that under this stricter condition, classical strategies face fundamental limitations--yet quantum entanglement enables players to overcome them, reshaping the possibilities of faultless strategic decision-making under uncertainty.

Shannon’s noisy channel coding theorem establishes the channel capacity as the fundamental limit for reliable communication~\cite{Shannon1948}. For any discrete memory-less classical channel, one can construct encoding and decoding schemes that achieve arbitrarily low error probabilities in the asymptotic limit of many channel uses. A stricter notion of reliable communication is provided by \emph{zero‑error channel coding}, also introduced by Shannon~\cite{Shannon1956}. This setting considers the maximum rate at which messages can be transmitted such that the receiver never confuses one message for another—even in a single use of the channel. In contrast to standard channel coding, which tolerates a small error probability that vanishes asymptotically, zero‑error coding requires perfect decoding with complete certainty on every use. Despite its conceptual clarity, the problem of determining the zero-error capacity of a general classical channel is {non-computable}~\cite{Boche2020, Boche2020b}. Interestingly, later studies have shown that {quantum entanglement} can boost this capacity well beyond classical limits~\cite{Cubitt2010, Cubitt2011, Yadavalli2022, Mir20203, Agarwal2024}.

To explore the implications of zero-error {performance} in strategic decision-making, we begin with the framework of game theory. A central concept in this framework is the Nash equilibrium \cite{Nash1950, Nash1951}, 
{which describes a strategy profile in which no player can increase their payoff / utility by unilaterally deviating from their chosen action.}
Nash showed that every finite game always admits at least one mixed strategy equilibrium, assuming players have complete knowledge of the game structure and act independently. This notion was later extended to situations involving uncertainty about other players’ types, each being a parameter representing private information relevant to strategic choices.

{Later, Harsanyi introduced the model of Bayesian games \cite{Harsanyi1967i,Harsanyi1968ii,Harsanyi1968iii}, in which each player is assigned a type drawn from a probability distribution. Players do not know each other’s types and therefore choose strategies based on beliefs over those types, seeking to maximise expected payoff. In this setting, several equilibrium concepts extend the notion of Nash equilibrium by averaging payoffs across type profiles — including correlated equilibrium, social-welfare solutions. In what follows, we focus specifically on the correlated equilibrium as a natural framework for analysing coordination under uncertainty.}

In Nash equilibrium, players choose their strategies independently. In contrast, Aumann introduced the idea of correlated equilibrium  \cite{Aumann1974,Aumann1987}. In this setting, players may make decisions based on shared correlation that comes from a common source. Aumann showed that if all players are Bayesian rational, that is, each maximizes expected utility using private information and a common prior and if this is common knowledge \cite{Aumann1976}, then their joint actions form a correlated equilibrium. Aumann's result shows that correlated equilibrium is a mathematically more general solution concept than Nash equilibrium: while every Nash equilibrium induces a correlated equilibrium, the converse does not hold. The set of all correlated equilibria forms a convex polytope defined by linear constraints, and every Nash equilibrium lies within this set. As a result, some of the strategic outcomes induced by correlated equilibria cannot be reached through uncorrelated or independent play.

The idea of shared correlation as a resource for realizing correlated equilibria in game-theoretic settings highlights a natural synergy with a foundational question from physics: what kinds of correlations are physically possible, and how do they differ from classical expectations? While this question lies squarely within the domain of physics, game theory stands to benefit from its resolution. Bell’s theorem \cite{Bell1964,Brunner2014} offers a sharp insight into this question, demonstrating that certain correlations observed in nature cannot be explained by any classical framework based on local realism. Instead, quantum mechanics—with its distinctive features such as entanglement and measurement incompatibility—offers access to these non-classical correlations \cite{Aspect1982,Hensen2015,Giustina2015,Shalm2015}. From this perspective, it becomes compelling to explore how such quantum resources can be meaningfully incorporated into game-theoretic frameworks, offering new operational insights that are relevant to both disciplines.

The incorporation of quantum correlations into game theoretic models enables new forms of strategic coordination beyond classical limits. When players are equipped with quantum resources they can implement strategies that are otherwise infeasible. One of the earliest explorations in this direction was carried out by Eisert et al. \cite{Eisert1999}, who proposed a quantized version of the Prisoner’s Dilemma. They showed that the classical dilemma can disappear if quantum strategies are allowed, yielding a quantum advantage over classical strategies. This initiated a sustained line of inquiry into quantum game theory \cite{Meyer1999,Flitney2003}, which gradually expanded into more general settings. In particular, later works explored scenarios where Bell nonlocal correlations allow players to achieve outcomes that are unattainable by classical strategies~\cite{Brunner2013,Pappa2015}. In this later phase of research, particularly in works exploring Bell nonlocality, the underlying game theoretic structure was more general than the standard strategic form games and was often modeled as correlated quantum equilibirum in Bayesian games with incomplete information. A series of recent results has further highlighted the advantages offered by quantum correlations in such settings~\cite{Roy2016,Banik2019,Rai2017,Auletta2021,Groisman2020,Abbott2024,BolonekLaso2017,Moreno2020}. {These works often demonstrate quantum advantage in social welfare solutions which maximize the sum of players’ expected payoffs, where each expectation is taken over all type profiles.}

{Building on this foundation, we introduce a new class of Bayesian games in which strict constraints are imposed on players’ actions or outcomes at every type profile.} Specifically, we consider a {Bayesian game} setting in which two spatially separated, non-communicating players receive private inputs (or types) drawn from a shared prior distribution. {Each type pair induces a corresponding payoff matrix for the players’ actions in that round, and this may admit multiple Nash equilibria.} {Since players in a Bayesian game know only their own type, they are generally unaware of which payoff matrix is in effect in a given round of play.}

Within this framework, we impose a particularly stringent winning condition: a strategy is deemed successful {if} and only if it guarantees that the players' chosen actions constitute a Nash equilibrium of the appropriate payoff matrix in \emph{every} round, regardless of the realized type pair. 
{We refer to this requirement as achieving a \textit{zero-error Nash equilibrium} -- a condition especially relevant in high-stakes strategic settings where errors are intolerable. A well-known illustration is the Game of Chicken \cite{Osborne}, invoked by Bertrand Russell in the context of nuclear brinkmanship \cite{Russell1959}. While this is not strictly a Bayesian game, it underscores how certain scenarios admit no margin for error. In the Bayesian setting, where players must act under uncertainty about others’ private information, the zero-error requirement becomes even more intricate but also more pertinent to real-world situations in which coordination must succeed despite incomplete knowledge.}
{The term “zero-error” is borrowed from Shannon’s communication paradigm, where flawless decoding is demanded; here, we adapt the idea to strategic decision making under type uncertainty.}

{We present several examples of Bayesian games in which classical resources are insufficient to meet the zero-error Nash equilibrium condition. {However}, quantum resources, specifically, shared entangled states offer a decisive advantage. Just as classical correlated equilibria rely on shared randomness or common advice to coordinate players' strategies, here some entanglement functions as a non-classical form of shared correlation or advice. With access to such entangled correlations, players can coordinate their actions in a way that ensures a zero-error Nash equilibrium is achieved for every possible type realization. We refer to this phenomenon as a {quantum advantage in zero-error Bayesian games}.} 

{In this work, we demonstrate this advantage through three representative constructions. We begin with a two-player Bayesian game where classical strategies inevitably fail, yet an entanglement-assisted protocol achieves flawless coordination. We then extend the framework to a three-player setting, showing that genuine multipartite entanglement enables zero-error equilibrium coordination unattainable with any classical resource. Finally, we consider a minimal two-player game with only two types and two actions per player, under a strengthened zero-error requirement, and establish that even here quantum correlations succeed—remarkably, every two-qubit pure entangled state except the maximally entangled one, when used as common advice between the players, provides the needed advantage. These results collectively illustrate both the breadth and the subtlety of quantum resources in zero-error Bayesian games, setting the stage for the detailed analyses that follow. Beyond these idealized settings, we further show that the quantum advantage persists under realistic noise, yielding near-zero error coordination that outperforms all classical strategies up to a certain noise threshold.}

{The remainder of the paper is organized as follows. Section \ref{framework} develops the formal framework of zero-error Nash equilibrium in Bayesian games, from simple examples to a precise definition. Section \ref{2pl} presents a two-player game where classical strategies fail but quantum entanglement achieves flawless coordination. Section \ref{3pl} extends this to three players, requiring genuine multipartite entanglement. Section \ref{minimal} introduces a minimal two-player game with a strengthened zero-error requirement, where non-maximal entanglement suffices. Section \ref{noise} analyzes coordination robustness under depolarizing noise, and Section \ref{disc} concludes with broader implications.}


\section{Framework}\label{framework}
We now turn to the formal framework in which the notion of zero-error Nash equilibrium can be precisely formulated. In the previous section, we introduced this concept as a coordination requirement inspired by zero-error communication, where players must ensure that their chosen strategies yield a Nash equilibrium for every type realization without any probability of failure. This requirement becomes especially relevant in Bayesian games, where players hold private information and lack communication. Our objective in this subsection is to define and understand zero-error Nash equilibrium within this classical setting, before turning to quantum strategies in the next part of the paper.

We proceed by presenting a sequence of illustrative examples that clarify the operational meaning of zero-error Nash equilibrium and demonstrate how such solutions can or cannot be achieved using classical strategies alone. The idea is developed incrementally through a sequence of examples, culminating in a formal definition that captures the zero-error constraint in a precise mathematical setting.

We first recall the formal definition of a Nash equilibrium in finite games \cite{Osborne}.  Consider a setting involving \( n \) players, where each player \( i \in \{1, 2, \dots, n\} \) chooses an action from a finite set \( A_i = \{a_i^1, a_i^2, \dots, a_i^{k_i}\} \). The utility of player \( i \) is given by a function
$u_i : A_1 \times A_2 \times \cdots \times A_n \rightarrow \mathbb{R}$,
which maps joint action profiles to real-valued payoffs.

A \emph{Nash equilibrium} is a strategy profile \( (a_1^*, a_2^*, \dots, a_n^*) \in A_1 \times A_2 \times \cdots \times A_n \) such that, for every player \( i \),
\[
u_i(a_i^*, a_{-i}^*) \geq u_i(a_i, a_{-i}^*) \quad \text{for all } a_i \in A_i,
\]
where \( a_{-i}^* \) denotes the strategies of all players except \( i \). That is, no player can unilaterally deviate and increase their utility. The above definition is actually of a pure strategy Nash equilibrium. There are concepts of mixed strategy Nash equilibrium as well. However, throughout our discussion, we will only talk about games with pure strategy Nash equilibria. 

We illustrate this with a simple two-player game \( G_1 \), involving players Alice and Bob. Each selects an action \( (a, b) \in A \times B = \{0,1\} \times \{0,1\} \), and the utilities are defined as:
\[
(u_A(a,b), u_B(a,b)) = 
\begin{cases}
(1,1) & \text{if } a = 0 \text{ and } b = 0, \\
(0,0) & \text{otherwise}.
\end{cases}
\]
The unique Nash equilibrium is \( (a,b) = (0,0) \), which the players can trivially coordinate on, even without communication.

Now consider a variation \( G_2 \), in which both \( (a,b) = (0,0) \) and \( (1,1) \) yield utility \( (1,1) \). This game admits two pure-strategy Nash equilibria. Without coordination, the players may select mismatched actions and fail to reach an equilibrium. However, if they share access to common randomness---such as perfectly correlated coin flips---they can always coordinate on one of the equilibria, achieving a \emph{zero-error Nash equilibrium}. {While $G_1$ and $G_2$ involve no private inputs or uncertainty, they can be viewed as special cases of a broader framework: Bayesian games \cite{Osborne}.}

{A two-player Bayesian game \( G = (X, Y, \mu,A,B, u_A^{xy}, u_B^{xy}) \) where \( X \) and \( Y \) are the players’ type spaces, \( \mu(x,y) \) is the prior distribution over types, $A,B$ are the strategy sets of the players and \( u_A^{xy}:A\times B \rightarrow \mathbb{R}, u_B^{xy}:A\times B \rightarrow \mathbb{R} \) define the utility functions for each type pair \( (x,y) \). In each round, players receive private types drawn from $\mu$, and their objective is to choose actions that maximize their payoffs. When the type pair is fixed and commonly known, the setting reduces to an ordinary game, such as $G_1$ or $G_2$. }

{To illustrate the genuinely Bayesian case,} we define a game \( G_3 \) in which Alice and Bob receive types or private inputs \( (x, y) \in X \times Y = \{0,1\} \times \{0,1\} \), drawn from a shared prior distribution---e.g., the uniform distribution over \( X \times Y \). For each type pair \( (x, y) \), the corresponding utility functions \( u_A^{xy} \) and \( u_B^{xy} \) are defined by:
\[
(u_A^{xy}(a,b), u_B^{xy}(a,b)) = 
\begin{cases}
(1,1) & \text{if } a = x \text{ and } b = y, \\
(0,0) & \text{otherwise}.
\end{cases}
\]
Each payoff matrix \( (u_A^{xy}, u_B^{xy}) \) admits a unique pure-strategy Nash equilibrium at \( (a,b) = (x,y) \). Despite having access only to their own private type, each player can deterministically choose their action: Alice sets \( a = x \), Bob sets \( b = y \), thus ensuring that the equilibrium is always achieved. In this case, a zero-error Nash equilibrium can be attained without the need for any shared randomness or quantum resources.

{Now consider a little more complex Bayesian game $G_4$. As before, the players receive types $(x,y)\in\{0,1\}^2$, drawn from a shared prior. The utilities depend on the relation between the types: for $(x,y)\in\{(0,0),(1,1)\}$, the players obtain payoff $(1,1)$ if their outputs satisfy $a=b$, and $(0,0)$ otherwise; for $(x,y)\in\{(0,1),(1,0)\}$, they obtain payoff $(1,1)$ if $a\neq b$ and $(0,0)$ otherwise. This game admits zero-error Nash equilibria: the players simply play $a=x,\,b=y$, which always yields a valid equilibrium outcome. Note that shared randomness does not expand the set of such perfect strategies, any mixture of deterministic strategies that succeeds with certainty must already contain a deterministic one that does so. Thus shared randomness brings no additional advantage.}

{We now formalize the notion of a zero-error Nash equilibrium within the Bayesian game framework. We say that a game} \emph{admits a zero-error Nash equilibrium} if there exists a shared resource that is, a random variable with a finite support set \( \Lambda \) and a known distribution \( p(\lambda) \), and a pair of local response functions:
$f : X \times \Lambda \rightarrow A, \quad g : Y \times \Lambda \rightarrow B$,
such that for every \( (x, y, \lambda) \), the \textit{output profile} \( (f(x, \lambda), g(y, \lambda)) \) is a {Nash equilibrium} of the game specified by \( (u_A^{xy}, u_B^{xy}) \).

{In other words, a {zero-error Nash equilibrium} exists in a Bayesian game if the players can always (i.e., with zero error) produce an action profile that constitutes a Nash equilibrium of the appropriate game for every realized type pair---possibly using shared correlations. Importantly, this notion does not extend or alter the definition of Nash equilibrium itself; it does not introduce a new equilibrium concept—such as correlated or Pareto-optimal equilibrium. Rather, it imposes a realizability constraint: the players must be able to coordinate on a valid Nash equilibrium of the underlying stage game for every type realization, without error. Thus, it serves as a criterion for faultless strategic decision-making under uncertainty.}

\begin{table}[]
    \centering
    \[
\begin{array}{c|c|c|c|}
     & y_1 & y_2 & y_3 \\
\hline
x_1 & \small\shortstack{\vspace{2pt} \\ 1,1 / 1,2 / 2,1 / 2,2 \\ 3,3 / 3,4 / 4,3 / 4,4} &   \small\shortstack{\vspace{2pt} \\ 1,1 / 1,2 / 2,3 / 2,4 \\ 3,1 / 3,2 / 4,3 / 4,4}  &    \small\shortstack{\vspace{2pt} \\ 1,1 / 1,2 / 2,3 / 2,4 \\ 3,3 / 3,4 / 4,1 / 4,2}  \\
\hline
x_2 &  \small\shortstack{\vspace{2pt} \\ 1,1 / 1,3 / 2,1 / 2,3 \\ 3,2 / 3,4 / 4,2 / 4,4}   &  \small\shortstack{\vspace{2pt} \\ 1,1 / 1,3 / 2,2 / 2,4 \\ 3,1 / 3,3 / 4,2 / 4,4}    &   \small\shortstack{\vspace{2pt} \\ 1,1 / 1,3 / 2,2 / 2,4 \\ 3,2 / 3,4 / 4,1 / 4,3}   \\
\hline
x_3 &  \small\shortstack{\vspace{2pt} \\ 1,1 / 1,4 / 2,1 / 2,4 \\ 3,2 / 3,3 / 4,2 / 4,3}    &  \small\shortstack{\vspace{2pt} \\ 1,1 / 1,4 / 2,2 / 2,3 \\ 3,1 / 3,4 / 4,2 / 4,3}    &  \small\shortstack{\vspace{2pt} \\ 1,2 / 1,3 / 2,1 / 2,4 \\ 3,1 / 3,4 / 4,2 / 4,3}    \\
\hline
\end{array}
\]
\caption{{The Bayesian game $G_5$, with private type sets 
$X = \{x_1, x_2, x_3\}$ for Alice and $Y = \{y_1, y_2, y_3\}$ for Bob. 
Each cell corresponding to a type pair $(x_i, y_j)$ lists the set 
$S_{ij} \subseteq A \times B$ of joint actions that constitute Nash equilibria 
of the stage game for that type pair, where $A = B = \{1,2,3,4\}$. 
An action pair yields payoff~1 if and only if it belongs to $S_{ij}$, and~0 otherwise. 
The coordination task is to select, for each type pair, a valid equilibrium action 
without communication. This is precisely the \emph{zero-error Nash equilibrium} coordination problem. 
No classical strategy can guarantee success, while a quantum strategy using entanglement as shared advice does.}}

    \label{tab:my_label}
\end{table}

\section{{Quantum Advantage in a Two-Player Bayesian Game}} \label{2pl}

{{We now turn to a Bayesian game that reveals a quantum advantage in coordination capabilities.} To motivate the setting, consider a realistic scenario from cybersecurity, where exact decision alignment under uncertainty is essential.}

{Two cybersecurity analysts, Alice and Bob, are responsible for independently configuring defense protocols at two remote data centers. Each site is subject to different classes of threats: internal anomalies for Alice and external network activity for Bob. After independently analyzing local system data, Alice receives a private alert \( x \in \{x_1, x_2, x_3\} \), representing increasing severity of internal compromise, while Bob receives a private report \( y \in \{y_1, y_2, y_3\} \) reflecting the external threat level. These inputs are sampled independently and remain private—each analyst is unaware of the other's type due to security reasons.}

{Each analyst must choose a defense configuration from a fixed set of four {different} protocols, encoded as actions \( a \in A = \{1,2,3,4\} \) and \( b \in B = \{1,2,3,4\} \). The success of the overall system depends on whether the two selected configurations are compatible for the threat conditions \( (x_i, y_j) \). These compatibility constraints are established in advance and are based on organizational security policies and inter-system requirements.}

{Formally, we model this situation as a two-player Bayesian game \( G_5 \). Each player receives a private type from the sets \( X = \{x_1, x_2, x_3\} \) and \( Y = \{y_1, y_2, y_3\} \), and chooses an action from their respective action sets \( A = \{1,2,3,4\} \) and \( B = \{1,2,3,4\} \). The structure of the game is encoded in Table~\ref{tab:my_label}, where each cell corresponds to a specific type pair \( (x_i, y_j) \in X \times Y \), and lists a subset \( S_{ij} \subseteq A \times B \) consisting of {eight} \textit{allowed} action pairs.}

{These allowed action pairs have a game-theoretic interpretation: each type pair \( (x_i, y_j) \) defines a distinct stage game whose payoff function assigns utility 1 to the action pairs in \( S_{ij} \) and utility 0 to all others. Thus, each \( S_{ij} \) corresponds to the set of \textit{pure-strategy Nash equilibria} of the stage game induced by the realized type profile. That is, for any \( (a, b) \in S_{ij} \), neither player has an incentive to unilaterally deviate from their action, given their private type and the underlying game structure. The task of the players is to coordinate—without communication—on one of these equilibrium action pairs for each realized type profile.}

{This requirement defines a stringent coordination condition: a valid strategy must guarantee that for every possible realization \( (x_i, y_j) \), the joint action \( (a_i, b_j) \) lies in the corresponding equilibrium set \( S_{ij} \). Any deviation from this constraint results in zero payoff for both players, and thus constitutes a failure of coordination. We refer to such a strategy, which guarantees equilibrium alignment for all type realizations, as achieving a zero-error Nash equilibrium in the Bayesian game.}

{In what follows, we demonstrate that no classical strategy can satisfy this constraint in \( G_5 \). Every such strategy will fail on at least one type profile. 
{However, if the players are given shared quantum advice, in the form of an entangled state they can each perform a local measurement that depends only on their private information. The outcomes of these measurements then guide their choices of actions. This procedure, called a quantum strategy \cite{Brunner2013,Pappa2015,Roy2016,Banik2019}, ensures that the players always arrive at a valid equilibrium outcome, something that cannot generally be guaranteed with classical shared advice.}
This establishes a robust quantum advantage in the task of error-free equilibrium coordination under private information.}

\subsection{Classical Limitations: Local Realism and Its Failure}

{The impossibility of achieving zero-error coordination in the game \( G_5 \) using classical strategies prompts a deeper question: what exactly are the constraints imposed by classical models of correlation? To formalize this, we draw on a foundational concept from physics—\emph{local realism}—which underlies all classical descriptions of distributed systems \cite{Bell1964,Brunner2014}.}

{In this framework, each player receives a private input, or \emph{type}, drawn from a known joint distribution. The players are allowed to share some initial classical correlation—represented by a shared random variable \( \lambda \), drawn from a probability distribution \( p(\lambda) \)—but are not allowed to communicate after receiving their types. Their outputs (i.e., their chosen actions) are determined by \emph{local deterministic response functions}, which map their private type and the shared variable to an action.}

{Formally, this means that any classical strategy must be described by a tuple $(\Lambda, p(\lambda), f, g)$,
where, \( \Lambda \) is a finite set of hidden variables, \( p(\lambda) \) is a probability distribution over \( \Lambda \), \( f : X \times \Lambda \to A \) is Alice’s local response function, and \( g : Y \times \Lambda \to B \) is Bob’s local response function.}

{Given such a model, the joint output distribution is given by:
\[
P(a,b \mid x_i,y_j) = \sum_{\lambda \in \Lambda} p(\lambda)\; \delta_{a, f(x_i,\lambda)}\; \delta_{b, g(y_j,\lambda)},
\]
where \( \delta \) denotes the Kronecker delta function.}
{This expression captures the full range of classical correlations achievable in the no-communication setting.}

{In the context of zero-error Bayesian games, this implies that any classical strategy capable of achieving zero-error coordination across all type pairs must admit such a hidden-variable representation. That is, there exists some \( \lambda \in \Lambda \) for which the output pair of the input \( (x_i, y_j) \in X \times Y \), \textit{i.e.}  \( (f(x_i,\lambda), g(y_j,\lambda)) \) must lie within the set \( S_{ij} \subseteq A \times B \) of Nash equilibrium strategies for that type pair.}

{In what follows, we show that no such local hidden-variable model exists for the game \( G_5 \): there is \emph{no choice} of \( (\Lambda, p(\lambda), f, g) \) that satisfies this condition for all \( (x_i, y_j) \). This establishes a fundamental limitation of classical resources and paves the way for demonstrating a quantum advantage.}

{We now demonstrate that such a hidden-variable model cannot exist for the game \( G_5 \). Fix some hidden variable \( \lambda \in \Lambda \), and consider the deterministic functions \( f_\lambda(x) := f(x, \lambda) \) and \( g_\lambda(y) := g(y, \lambda) \). Assume, for instance, that \( f_\lambda(x_1) = 1 \).}

{From the table \( S_{11} \), this implies that \( g_\lambda(y_1) \in \{1,2\} \). Next, consider \( S_{12} \): since \( f_\lambda(x_1) = 1 \), it follows that \( g_\lambda(y_2) \in \{1,2\} \). Similarly, from \( S_{13} \), we obtain \( g_\lambda(y_3) \in \{1,2\} \). Let us suppose \( g_\lambda(y_3) = 1 \).}

{Now consider cell \( S_{23} \): for \( g_\lambda(y_3) =1 \), the possible values for \( f_\lambda(x_2) \) are \( \{1,4\} \). Let us analyze both branches:
\begin{itemize}
    \item If \( f_\lambda(x_2) = 1 \), then from \( S_{21} \) we must have \( g_\lambda(y_1) \in \{1,3\} \). But earlier we concluded \( g_\lambda(y_1) \in \{1,2\} \), so consistency requires \( g_\lambda(y_1) = 1 \).
    \item Otherwise, if \( f_\lambda(x_2) = 4 \), then from \( S_{21} \) we require \( g_\lambda(y_1) \in \{2,4\} \), implying \( g_\lambda(y_1) = 2 \). In either case, the value of \( g_\lambda(y_1) \) becomes fixed.
\end{itemize}
}

{Proceeding similarly for \( S_{22} \), we find that:
\begin{itemize}
    \item If \( f_\lambda(x_2) = 1 \), then \( g_\lambda(y_2) = 1 \);
    \item Otherwise, if \( f_\lambda(x_2) = 4 \), then \( g_\lambda(y_2) = 2 \).
\end{itemize}
}

{Now, consider cell \( S_{33} \): given \( g_\lambda(y_3) = 1 \), it follows that \( f_\lambda(x_3) \in \{2,3\} \). Next, from cell \( S_{31} \), we know:
\begin{itemize}
    \item If \( g_\lambda(y_1) = 1 \), then \( f_\lambda(x_3) \in \{1,2\} \), eventually, \( f_\lambda(x_3)=2\);
    \item If \( g_\lambda(y_1) = 2 \), then \( f_\lambda(x_3) \in \{3,4\} \) , eventually, \( f_\lambda(x_3)=3\).
\end{itemize}
}

{Likewise, from \( S_{32} \):
\begin{itemize}
    \item If \( g_\lambda(y_2) = 1 \), then \( f_\lambda(x_3) \in \{1,3\} \), , eventually, \( f_\lambda(x_3)=3\);
    \item If \( g_\lambda(y_2) = 2 \), then \( f_\lambda(x_3) \in \{2,4\} \) , eventually, \( f_\lambda(x_3)=2\).
\end{itemize}
}

{This shows a contradiction. The value of \( f_\lambda(x_3) \) depends on whether Bob’s input is \( y_1 \) or \( y_2 \), violating the locality assumption which requires Alice’s response function \( f_\lambda \) to be independent of Bob’s input.
}

{The same contradiction arises if we assume \( g_\lambda(y_3) = 2 \) instead of 1.}

{Finally, one can check that starting with any admissible value \( f_\lambda(x_1) \in \{1,2,3,4\} \) leads to analogous contradictions through similar dependency chains. Each branch of the tree eventually results in a condition where one player's output must depend on the other's input ruling out any LHV realization.}

{Therefore, no classical model—no choice of \( (\Lambda, p(\lambda), f, g) \) can generate responses that satisfy all Nash constraints encoded in the sets \( S_{ij} \) for every type pair \( (x_i, y_j) \in X \times Y \). This establishes that the game \( G_5 \) is not simulable by any local realistic hidden-variable model.}

\subsection{Quantum Strategy for Zero-Error Coordination}\label{quantum strategy}

{We now present a quantum strategy that achieves perfect, zero-error coordination in the game \( G_5 \), using two ebits of entanglement and carefully structured projective measurements. Alice and Bob share the entangled state \( \ket{\psi} = \ket{\phi^+}^{\otimes 2} = \left(\tfrac{1}{\sqrt{2}}(\ket{00} + \ket{11})\right)^{\otimes 2} \in \mathbb{C}^4 \otimes \mathbb{C}^4 \), which here serves as their \emph{quantum advice}, the nonclassical analogue of the shared advice used in Aumann's correlated equilibrium. {Upon receiving their respective 
private types \( x_i \in X \) and \( y_j \in Y \), each player performs a four-outcome 
projective measurement on their half of the shared state, specified by the projectors}
\[
\begin{aligned}
P_1 &= \tfrac{1}{4}(I + M + M' + MM'), \\
P_2 &= \tfrac{1}{4}(I + M - M' - MM'), \\
P_3 &= \tfrac{1}{4}(I - M + M' - MM'), \\
P_4 &= \tfrac{1}{4}(I - M - M' + MM'),
\end{aligned}
\]
{where $I$ is the identity operator on $\mathbb{C}^4$ and $M,M'$ are fixed Hermitian 
operators defining the measurement. The set $\{P_1,P_2,P_3,P_4\}$ constitutes a valid 
projective measurement on $\mathbb{C}^4$, applied locally by each player on their 
subsystem of $\mathbb{C}^4 \otimes \mathbb{C}^4$. We define \( M \) and \( M' \) as follows:} if Alice’s type is \( x_1 \), then \( M = Z \otimes I \) and \( M' = I \otimes Z \); if her type is \( x_2 \), then \( M = I \otimes X \) and \( M' = X \otimes I \); and if her type is \( x_3 \), then \( M = Z \otimes X \) and \( M' = X \otimes Z \). Bob selects his observables similarly: if his type is \( y_1 \), he uses \( M = Z \otimes I \) and \( M' = I \otimes X \); if \( y_2 \), then \( M = I \otimes Z \) and \( M' = X \otimes I \); and if \( y_3 \), then \( M = Z \otimes Z \) and \( M' = X \otimes X \). 
After performing their respective measurements, both players post-process their four-valued outcomes using a fixed, publicly known mapping into the output set \( \{1, 2,3,4\} \), with the mapping depending on the players’ types. For types \( x_1, x_2, x_3 \) and \( y_1, y_2,y_3 \), the outcomes \( P_1, P_2, P_3, P_4 \) are mapped to \( 1,2,3,4 \) respectively. {Consequently, for every type pair \( (x_i, y_j) \), the resulting joint output lies within the prescribed set \( S_{ij} \subseteq A \times B \).}
This constitutes an instance of a \emph{zero-error Nash equilibrium}. No classical strategy can satisfy this condition {in the game \( G_5 \).  The quantum strategy, }on the other hand, succeeds precisely where classical strategies fail. {This establishes a quantum advantage in the fault-tolerant decision-making framework, showing how entanglement-assisted strategies outperform all classical approaches.}}

\begin{figure}[t!]
\centering
\scriptsize
\begin{minipage}{0.48\columnwidth}
\centering
\renewcommand{\arraystretch}{1.2}
\resizebox{\linewidth}{!}{$
\begin{array}{c|c|c|}
z_1 & y_1 & y_2 \\
\hline
x_1 & \scriptsize\shortstack{\vspace{2pt} \\ 1,1,1 / 1,2,2 \\ 2,1,2 / 2,2,1} & \scriptsize\shortstack{\vspace{2pt} \\ All 8\\actions} \\
\hline
x_2 & \scriptsize\shortstack{\vspace{2pt} \\ All 8\\actions} & \scriptsize\shortstack{\vspace{2pt} \\ 1,1,2 / 1,2,1 \\ 2,1,1 / 2,2,2} \\
\hline
\end{array}
$}
\caption*{(a) Action sets for type profile \((x_i, y_j, z_1)\)}
\end{minipage}
\hfill
\begin{minipage}{0.48\columnwidth}
\centering
\renewcommand{\arraystretch}{1.2}
\resizebox{\linewidth}{!}{$
\begin{array}{c|c|c|}
z_2 & y_1 & y_2 \\
\hline
x_1 & \scriptsize\shortstack{\vspace{2pt} \\ All 8\\actions} & \scriptsize\shortstack{\vspace{2pt} \\ 1,1,2 / 1,2,1 \\ 2,1,1 / 2,2,2} \\
\hline
x_2 & \scriptsize\shortstack{\vspace{2pt} \\ 1,1,2 / 1,2,1 \\ 2,1,1 / 2,2,2} & \scriptsize\shortstack{\vspace{2pt} \\ All 8\\actions} \\
\hline
\end{array}
$}
\caption*{(b) Action sets for type profile \((x_i, y_j, z_2)\)}
\end{minipage}
\caption{Bayesian game \( G_6 \) involves three players—Alice, Bob, and Charlie—with type sets \( X = \{x_1, x_2\} \), \( Y = \{y_1, y_2\} \), and \( Z = \{z_1, z_2\} \), respectively. Each subtable corresponds to a fixed type of Charlie and lists the allowed joint actions \( (a, b, c) \in \{1,2\}^{\times 3} \) for every type pair \( (x_i, y_j) \) of Alice and Bob. The entries represent only those action triples that yield a payoff of 1 to all three players; all other joint actions result in zero payoff. Each listed action triple constitutes a pure-strategy Nash equilibrium in the corresponding stage game.}
\label{fig:3pl}
\end{figure}

\section{{Quantum Advantage in a Three-Player Bayesian Game}}\label{3pl}

We now extend our analysis to a three-player Bayesian game in which the constraint of zero-error Nash equilibrium cannot be satisfied by any classical strategy. In contrast, quantum strategies leveraging shared {tripartite genuine} entanglement can successfully meet this constraint. The game we analyze, denoted \(G_6\), is depicted in Figure~\ref{fig:3pl}.

In this scenario, three spatially separated players—Alice, Bob, and Charlie—are each assigned a private type drawn from their respective type sets: \(X = \{x_1, x_2\}\), \(Y = \{y_1, y_2\}\), and \(Z = \{z_1, z_2\}\). These type values are private and independently distributed; no player has access to the types received by the others.

Upon receiving their types, each player independently selects an action from their respective strategy sets \(A\), \(B\), and \(C\), where each set contains two possible actions labeled by integers \(k \in [2]\). The Bayesian game \(G_6\) is defined by a collection of utility functions that depend on the full type triple \((x_i, y_j, z_k)\).

Figure~\ref{fig:3pl} provides a compact representation of the structure of \(G_6\). Each subtable corresponds to a fixed type \(z_k\) for Charlie and displays the utility structure across all combinations of types \(x_i\) and \(y_j\) for Alice and Bob. For each type triple \((x_i, y_j, z_k)\), a subset \(S_{ijk} \subseteq A \times B \times C\) of action profiles is specified. If the players select an action profile belonging to \(S_{ijk}\), they each receive a payoff of 1; otherwise, the payoff is 0 for all players. These sets—explicitly listed in the corresponding cells of the figure—comprise all pure-strategy Nash equilibria of the stage game realized by the type triple \((x_i, y_j, z_k)\).

This game imposes a coordination task among three agents who only have access to their private types and any shared resource available prior to the game. The objective is to ensure that, for each realized type triple \((x_i, y_j, z_k)\), the players choose an action triple in \(S_{ijk}\), thereby ensuring that their joint action constitutes a pure-strategy Nash equilibrium of the corresponding stage game. {This requirement is precisely the zero-error Nash equilibrium condition.} As shown in the Appendix \ref{app_gen}, no classical strategy can guarantee such zero-error Nash equilibrium coordination across all type combinations. However, a quantum strategy using shared genuine entanglement {as advice} achieves perfect coordination restricted only to Nash equilibria, thereby satisfying the zero-error Nash equilibrium condition for the game \(G_6\).

The corresponding quantum strategy proceeds as follows. The players share a genuinely entangled three-qubit Greenberger–Horne–Zeilinger (GHZ) state, 
$\ket{g}_{ABC} = \tfrac{1}{\sqrt{2}}\left( \ket{000} + \ket{111} \right).$
Each player performs a projective measurement based on their received type: if a player receives type \(t_1\) for \(t \in \{x, y, z\}\), they measure in the \(X\)-basis; otherwise, they measure in the \(Y\)-basis. With this strategy, the players perfectly satisfy the zero-error Nash equilibrium constraint across all type profiles.

\section{{Minimal Bayesian Game with Strengthened Zero-Error Constraint}} \label{minimal}

\begin{table}[t]
    \centering
    
    \begin{tabular}{c|c|c|c|}
      
        & $y_1$ & $y_2$ \\
        \hline
        $x_1$ & $1,1~ / ~1,2~ / ~2,1~ / 
        ~2,2$ & $1,1~ / ~2,1~ / ~2,2$ \\
        \hline
        $x_2$ & $1,1~ / 
        ~1,2~ / ~2,2$ & $1,2~ / ~2,1~ / 
        ~2,2$ \\
        \hline
    \end{tabular}
    \caption{{Bayesian game \( G_7 \) features minimal cardinalities. The row indices \( x_1, x_2 \) and column indices \( y_1, y_2 \) represent the private types of the two players. Each cell lists the allowed joint actions \( (a, b) \) for the corresponding type profile, where \( a, b \in \{1,2\} \) denote the actions chosen by the players, Alice and Bob respectively. The table includes only those action pairs that yield a payoff of 1 to both players; all omitted pairs result in zero payoff. Notably, every displayed action pair constitutes a Nash equilibrium in its respective stage game.
}}
\label{tab:Hardy's_table}
\end{table}

Thus far, we have examined coordination scenarios where quantum resources offer a clear advantage under the {zero-error Nash equilibrium} constraint. These games involved either a larger type space or multiple players. A natural direction is to investigate whether such {quantum advantage} persists in minimal Bayesian game settings — specifically, games involving only two players, each with two types and two actions.

{We now present such a game $G_7$ shown in Table \ref{tab:Hardy's_table}, constructed to operate within this minimal structure.} To demonstrate quantum advantage in this context, we adopt a {stronger version of the {zero-error Nash equilibrium} constraint}, requiring that: (i) all non-equilibrium outcomes occur with {zero probability}, and (ii) all equilibrium outcomes occur with {nonzero probability}.

No classical strategy can satisfy both these requirements. The {impossibility of any local hidden variable strategy} fulfilling this condition is formally shown in {Appendix~\ref{appendix:Hardy-proof}}.
{In contrast, quantum strategies leveraging shared entanglement can satisfy the zero-error coordination condition. Importantly, the quantum resource required is neither high-dimensional nor multipartite: any non-maximal two-qubit pure entangled state suffices, including those arbitrarily close to a product state. Appendix~\ref{app_hardyq} details the corresponding quantum advice and strategies. This highlights that non-maximal entanglement can sometimes be more effective than maximal entanglement in satisfying fine-grained coordination requirements.}

\section{{Noise Robustness and Near-Zero Error Coordination}}\label{noise}

{Thus far, our analysis has focused on the \textit{strict} zero-error requirement, where quantum strategies achieve coordination impossible for any classical counterpart. These results establish a sharp quantum advantage in the ideal, noise-free regime. In realistic settings, however, noise is unavoidable and renders exact satisfaction of the zero-error Nash equilibrium condition unattainable. The relevant question then is whether quantum resources can sustain an advantage under noisy conditions by achieving \textit{near-zero error} coordination—that is, maintaining a strictly smaller error probability than any classical strategy within a nonzero noise threshold.}

\begin{figure}[t!]
\centering
\includegraphics[scale=0.59]{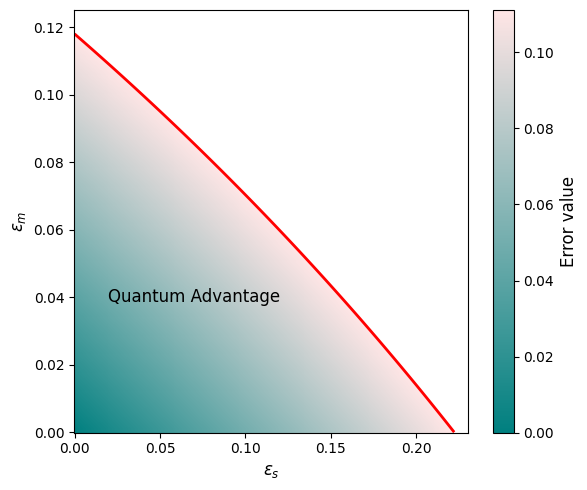}
\caption{{This plot shows the error probability as a gradient map, with colors ranging from green (zero error) to red (maximum error). It illustrates the trade-off between the noise parameters $\epsilon_s$ and $\epsilon_m$, and identifies the region in which quantum advantage is preserved in the $G_5$ game despite depolarizing noise affecting both the shared state and local measurements. The red boundary marks the threshold beyond which the quantum strategy ceases to outperform all classical ones.
}}
\label{fig1}
\vspace{-.5cm}
\end{figure}

{In this section, we examine robustness in the setting of the game $G_5$. 
As established in Sec.~\ref{quantum strategy}, perfect coordination requires Alice and Bob to share the entangled state $\ket{\psi} = \ket{\phi^+}^{\otimes 2}$. In practice, however, the source distributing entanglement is subject to imperfections and introduces depolarizing noise. For a depolarizing channel $\mathcal{D}_\epsilon$ with parameter $\epsilon \in [0,1]$, the action on any state $\rho$ is given by $\mathcal{D}_\epsilon(\rho) = (1-\epsilon)\rho + \frac{\epsilon}{16}\, I_{16}.$ Accordingly, the ideal resource transforms as
\begin{equation}
    \mathcal{D}_\epsilon\!\left(\ket{\phi^+}\!\bra{\phi^+}^{\otimes 2}\right) 
    = (1-\epsilon)\ket{\phi^+}\!\bra{\phi^+}^{\otimes 2} + \frac{\epsilon}{16}\, I_{16}.\nonumber
\end{equation}
}

{Noise may also affect the measurement devices of Alice and Bob. 
Under the same depolarizing model, a projector $P_1$ evolves as
\begin{equation}
    \mathcal{D}_\epsilon(P_1) = (1-\epsilon)P_1 + \frac{\epsilon}{4} I_4.\nonumber
\end{equation}
Let the shared state and local measurements be corrupted by channels $\mathcal{D}_{\epsilon_s}$ and $\mathcal{D}_{\epsilon_m}$, respectively. 
Then the probability that the outcome $(a,b)$ falls outside the allowed set $S_{ij}$ is
\begin{equation}
    P\!\left((a,b)\notin S_{ij}\,\middle|\,x_i,y_j\right) 
    = \frac{1}{2}\!\left[\,\epsilon_s(\epsilon_m-1)^2-(\epsilon_m-2)\epsilon_m\,\right]. \nonumber
\end{equation}
}
{For any classical deterministic strategy, the error probability is bounded below by $1/9$. 
Hence the quantum strategy retains an advantage whenever
\begin{equation}\epsilon_s(\epsilon_m-1)^2-(\epsilon_m-2)\epsilon_m < \tfrac{2}{9},\nonumber
\end{equation}
as depicted in Fig.~\ref{fig1}.
}

{Our analysis identifies the region of noise parameters $(\epsilon_s,\epsilon_m)$ where quantum strategies retain an advantage by achieving strictly smaller error probabilities than any classical counterpart.
In the absence of noise, this corresponds to satisfaction of the zero-error Nash equilibrium condition, whereas under finite noise it reduces to near-zero error coordination toward the same equilibrium. These findings highlight the robustness of the quantum advantage and delineate the operational window in which it can be harnessed on noisy intermediate-scale quantum platforms.}

\section{Discussion} \label{disc}

{Our work brings together three domains that have traditionally developed in isolation. The first is the zero-error paradigm from communication theory \cite{Shannon1956}, which formalizes the stringent requirement of coordination without any possibility of mistakes. The second is Bayesian game theory \cite{Harsanyi1967i,Harsanyi1968ii,Harsanyi1968iii}, where strategic agents must act under incomplete information. The third is the study of Bell nonlocal correlations \cite{Bell1964}, which, when viewed as shared advice, allow players to surpass the limits imposed by local realism. By embedding the zero-error condition into Bayesian games, we uncover a conceptual bridge between information theory and game theory: coordination becomes not only a matter of optimizing payoffs but of guaranteeing correctness. Crucially, we find Bayesian games where this requirement cannot be satisfied using classical correlations alone. In such cases, quantum correlations exhibiting Bell nonlocality provide the additional resource needed for players to achieve zero-error coordination. This highlights Bell nonlocality as the distinguishing feature that extends fault-tolerant coordination beyond what is possible in any classical framework.}

{Our game constructions are motivated by paradigmatic nonlocality arguments such as the Peres–Mermin magic square~\cite{Mermin1990,Peres1990,Aravind2002}, the GHZ paradox~\cite{GHZ1989}, and Hardy’s argument~\cite{Hardy1992,Hardy1993}, and are developed within a game-theoretic framework enforcing strict fault-tolerant equilibrium constraints. While these arguments expose \textit{sharp} conflicts with classical realism, the idealized correlations they rely upon remain difficult to achieve in practice. To address this, we introduce the role of noise and analyze its impact on zero-error coordination. Remarkably, we find that quantum strategies continue to outperform all classical ones within a finite regime of noise parameters. We refer to this phenomenon as near-zero error coordination.}

{Several open problems naturally emerge. A central direction concerns the characterization of quantum strategies that achieve zero-error Nash equilibria. Beyond this, one may ask whether the very notion of equilibrium refinements in game theory should be rethought in light of these nonclassical resources, suggesting a broader re-examination of rationality under uncertainty.}

\begin{acknowledgments}
We acknowledge the National Symposium on Quantum Information and Foundations (NSQIF2025), organized by the Indian Statistical Institute, Kolkata, where the seed of this work was first sown. We thank Manik Banik for pointing out the connection between the fault-tolerant attainability of Nash equilibrium studied here and Shannon’s zero-error notion in communication theory. We are also grateful to Guruprasad Kar, Tamal Guha, Subhendu B. Ghosh, Snehasish Roy Chowdhury for insightful discussions at various stages of this work. S.R.P. acknowledges the MEXT program. A.D.B. acknowledges STARS (STARS/STARS2/2023-0809), Govt. of India. A.M. acknowledges Research Initiation Grant provided by IIT Jodhpur. 
\end{acknowledgments}

\appendix

\section{Proof of no classical strategy for Game \(G_6\)}\label{app_gen}

Assume a local hidden‑variable model with shared randomness \(\lambda\) and deterministic response functions
\[
a = f(x_i,\lambda),\quad
b = g(y_j,\lambda),\quad
c = h(z_k,\lambda).
\]
Fix \(\lambda\) and write simply \(f(x_i)\), \(g(y_j)\), \(h(z_k)\).  For every \((x_i,y_j,z_k)\) the triple \(\bigl(f(x_i),g(y_j),h(z_k)\bigr)\) must lie in the Nash equilibrium set \(S_{ijk}\) from Fig. \ref{fig:3pl}.

\medskip
\noindent\textbf{Step 1:} Fix \(f(x_1)=1\).  Look at \((x_1,y_1,z_1)\); from Fig. \ref{fig:3pl}(a)
\[
S_{1,1,1} = \{(1,1,1),\,(1,2,2)\}.
\]
Hence either

\noindent\textbf{Case A: }$g(y_1)=1,\;h(z_1)=1$,
{or}\\
\textbf{Case B: }$g(y_1)=2,\;h(z_1)=2$.

\paragraph{Case A: \(g(y_1)=1,\;h(z_1)=1\).}
Now inspect \((x_1,y_2,z_2)\) (Fig. \ref{fig:3pl}(b)):
\[
S_{1,2,2} = \{(1,1,2),\,(1,2,1),\,(2,1,1),\,(2,2,2)\}.
\]
Since \(f(x_1)=1\), we must have \((g(y_2),h(z_2))\) equal to either \((1,2)\) or \((2,1)\).

\begin{itemize}
  \item \textbf{Subcase A1:} \(g(y_2)=1,\;h(z_2)=2\).  
    At \((x_2,y_2,z_1)\) (Fig. \ref{fig:3pl}(a)), the same four triples are allowed.  With \((g,h)=(1,1)\), only \((2,1,1)\) fits, so \(f(x_2)=2\).  
    But at \((x_2,y_1,z_2)\) (Fig. \ref{fig:3pl}(b)), those four allowed and \((g,h)=(1,2)\) force \((1,1,2)\), so \(f(x_2)=1\).  
    Contradiction: \(1\neq2\).

  \item \textbf{Subcase A2:} \(g(y_2)=2,\;h(z_2)=1\).  
    At \((x_2,y_2,z_1)\), \((g,h)=(2,1)\) forces \((1,2,1)\), so \(f(x_2)=1\).  
    Then at \((x_2,y_1,z_2)\), \((f,g,h)=(1,1,1)\) is not among the four allowed.  
    Contradiction.
\end{itemize}

\paragraph{Case B: \(g(y_1)=2,\;h(z_1)=2\).}
Again at \((x_1,y_2,z_2)\), since \(f(x_1)=1\), we must have \((g(y_2),h(z_2))=(1,2)\) or \((2,1)\).

\begin{itemize}
  \item \textbf{Subcase B1:} \(g(y_2)=1,\;h(z_2)=2\).  
    At \((x_2,y_2,z_1)\), \((g,h)=(1,2)\) forces \((1,1,2)\), so \(f(x_2)=1\).  
    At \((x_2,y_1,z_2)\), \((g,h)=(2,2)\) forces \((2,2,2)\), so \(f(x_2)=2\).  
    Contradiction.

  \item \textbf{Subcase B2:} \(g(y_2)=2,\;h(z_2)=1\).  
    At \((x_2,y_2,z_1)\), \((g,h)=(2,2)\) forces \((2,2,2)\), so \(f(x_2)=2\).  
    At \((x_2,y_1,z_2)\), \((g,h)=(2,1)\) forces \((1,2,1)\), so \(f(x_2)=1\). Contradiction.
\end{itemize}

\noindent\textbf{Step 2: \(f(x_1)=2\)}

Assume a local hidden‑variable model with shared \(\lambda\) and deterministic response functions
\[
a = f(x_i,\lambda),\quad b = g(y_j,\lambda),\quad c = h(z_k,\lambda).
\]
Fix \(\lambda\) and write \(f(x_i)\), \(g(y_j)\), \(h(z_k)\).  We now show that \emph{any} assignment with \(f(x_1)=2\) leads to a violation of the Nash‑equilibrium constraints in Fig. \ref{fig:3pl}.

\medskip
\noindent At \((x_1,y_1,z_1)\), Fig. \ref{fig:3pl}(a) gives
\[
S_{1,1,1} \;=\;\{(1,1,1),\,(1,2,2),\,(2,1,2),\,(2,2,1)\}.
\]
Since \(f(x_1)=2\), the only possible output triples are
\[
(2,1,2)\quad\text{or}\quad(2,2,1).
\]
Thus we must enter either
\begin{itemize}
  \item \textbf{Case A:} \((g(y_1),h(z_1))=(1,2)\), or
  \item \textbf{Case B:} \((g(y_1),h(z_1))=(2,1)\).
\end{itemize}

\medskip
\noindent\textbf{Case A: \(g(y_1)=1,\;h(z_1)=2\).}

\begin{enumerate}
  \item At \((x_1,y_2,z_2)\), Fig. \ref{fig:3pl}(b) gives
    \[
    S_{1,2,2} = \{(1,1,2),(1,2,1),(2,1,1),(2,2,2)\}.
    \]
    With \(f(x_1)=2\), the only candidates are
    \((2,1,1)\) or \((2,2,2)\).  Hence
    \((g(y_2),h(z_2)) = (1,1)\) or \((2,2)\).

  \item \emph{Subcase A1:} \(g(y_2)=1,\;h(z_2)=1\).  
    At \((x_2,y_2,z_1)\), the allowed four are again
    \(\{(1,1,2),(1,2,1),(2,1,1),(2,2,2)\}\).  Since 
    \((g,h)=(1,2)\), the only fit is \((1,1,2)\), so \(f(x_2)=1\).  
    But at \((x_2,y_1,z_2)\), the same four set with 
    \((g,h)=(1,1)\) forces \((2,1,1)\), so \(f(x_2)=2\).  
    Contradiction.

  \item \emph{Subcase A2:} \(g(y_2)=2,\;h(z_2)=2\).  
    At \((x_2,y_2,z_1)\), \((g,h)=(2,2)\) forces \((2,2,2)\), so \(f(x_2)=2\).  
    At \((x_2,y_1,z_2)\), \((g,h)=(1,2)\) forces $f(x_2)=1$.  
    Contradiction.
\end{enumerate}

\medskip
\noindent\textbf{Case B: \(g(y_1)=2,\;h(z_1)=1\).}

\begin{enumerate}
  \item At \((x_1,y_2,z_2)\), the same four
    \(\{(1,1,2),(1,2,1),(2,1,1),(2,2,2)\}\) appear.  With \(f(x_1)=2\),
    the only options are \((2,1,1)\) or \((2,2,2)\), so
    \((g(y_2),h(z_2)) = (1,1)\) or \((2,2)\).

  \item \emph{Subcase B1:} \(g(y_2)=1,\;h(z_2)=1\).  
    At \((x_2,y_2,z_1)\), \((g,h)=(1,1)\) forces \((2,1,1)\) implies \(f(x_2)=2\).  
    At \((x_2,y_1,z_2)\), \((g,h)=(2,1)\) forces \((1,2,1)\) implies \(f(x_2)=1\).  
    Contradiction.

  \item \emph{Subcase B2:} \(g(y_2)=2,\;h(z_2)=2\).  
    At \((x_2,y_2,z_1)\), \((g,h)=(2,1)\) forces \((1,2,1)\) implies \(f(x_2)=1\).  
    At \((x_2,y_1,z_2)\), \((g,h)=(2,2)\) forces \((2,2,2)\) implies \(f(x_2)=2\).  
    Contradiction.
\end{enumerate}

\bigskip
{Therefore no assignment with \(f(x_1)\) equals to $1$ or $2$ can satisfy all cells of Fig. \ref{fig:3pl}.} This completes the proof that no classical strategy exists for Game \(G_6\).


\section{Proof of no classical strategy for the minimal cardinality game}
\label{appendix:Hardy-proof}

Assume a local hidden‑variable model with shared randomness \(\lambda\) and deterministic response functions
\[
a = f(x_i,\lambda),\quad
b = g(y_j,\lambda).
\]
Fix \(\lambda\) and write simply \(f(x_i)\), \(g(y_j)\). For every \((x_i,y_j)\), the pair \((f(x_i), g(y_j))\) must lie in the allowed action set \(S_{ij}\) from \ref{tab:Hardy's_table}.

\medskip
\noindent\textbf{Step 1:} Fix \(f(x_2)=1\), \(g(y_2)=2\). Consider cell \((x_2, y_1)\); from \ref{tab:Hardy's_table}:
\[
S_{21} = \{(1,1),\,(2,1),\,(2,2)\}.
\]
Since \(f(x_2) = 1\), the only compatible pair is \((1,1)\), hence:
\[
g(y_1) = 1.
\]

\medskip
\noindent\textbf{Step 2:} Now consider cell \((x_1, y_2)\); \ref{tab:Hardy's_table} gives:
\[
S_{12} = \{(1,1),\,(2,1),\,(2,2)\}.
\]
Since \(g(y_2) = 2\), the only compatible value for \(f(x_1)\) is \(2\), because:
\[
(2,2) \in S_{12}.
\]

So far, we have:
\[
f(x_1) = 2,\quad f(x_2) = 1,\quad g(y_1) = 1,\quad g(y_2) = 2.
\]

\medskip
\noindent\textbf{Step 3:} Check cell \((x_1, y_1)\); \ref{tab:Hardy's_table} gives:
\[
S_{11} = \{(1,1),\,(1,2),\,(2,2)\}.
\]
But our assignment gives \((f(x_1), g(y_1)) = (2,1)\), which is \emph{not} in the set.

\medskip
The choice \((f(x_2), g(y_2)) = (1,2)\) leads to a contradiction at \((x_1, y_1)\). Hence, this deterministic strategy cannot exist.

\medskip
A similar contradiction arises for all possible deterministic combinations. Thus:

{\text{No local hidden‑variable model exists for the game in \ref{tab:Hardy's_table}.}}

\section{Proof of quantum advantage in satisfying the stronger zero error Nash equilibrium condition in minimal cardinality game} \label{app_hardyq}

The strategy involves a two-qubit pure entangled state that is shared as advice between the players. Depending on their types, the players perform different measurements on their respective subsystems. For types $x_1$ and $x_2$, Alice performs the following measurements on her subsystem: an X-basis spin measurement and $\{\ket{a_0},\ket{a_1}\}$, respectively. For types $y_1$ and $y_2$, Bob performs an X-basis spin measurement and $\{\ket{b_0},\ket{b_1}\}$, respectively. 

Here, 
\[
\ket{a_0} = \cos \tfrac{\gamma}{2} \ket{0} + e^{\iota \eta} \sin \tfrac{\gamma}{2} \ket{1}, \quad \braket{a_0|a_1} = 0,
\]
and
\[
\ket{b_0} = \cos \tfrac{\delta}{2} \ket{0} + e^{\iota \kappa} \sin \tfrac{\delta}{2} \ket{1}, \quad \braket{b_0|b_1} = 0,
\]
where $\gamma,\delta \in [0,\pi]$ and $\eta,\kappa \in [0,2\pi)$. 

The entangled state shared between them is of the form
\[
\ket{\psi} = \frac{\left( \tan \tfrac{\gamma}{2} \ket{a_0 b_1} + \tan \tfrac{\delta}{2} \ket{a_1 b_0} + \ket{a_1 b_1} \right)}{\sqrt{1 + \tan^2 \tfrac{\gamma}{2} \tan^2 \tfrac{\delta}{2}}}.
\]

This quantum strategy, together with the shared quantum advice, leads to the satisfaction of the zero-error Nash equilibrium condition for the game $G_7$~\cite{Hardy1992,Hardy1993,Jordan1994,Kar1997,Rai2022}.





\end{document}